\documentstyle[11pt]{article}
\begin{document}

 \newcommand \be {\begin{equation}}
\newcommand \ee {\end{equation}}
 \newcommand \ba {\begin{eqnarray}}
\newcommand \ea {\end{eqnarray}}
\def\oppropto{\mathop{\propto}}
\def\operarrow{\mathop{\longrightarrow}}
\def\opsimeq{\mathop{\simeq}}

\title{\bf EXPONENTIAL FUNCTIONALS \\
OF BROWNIAN MOTION \\
AND DISORDERED SYSTEMS}

\vskip 3 true cm

\author{  Alain Comtet$^{(1)}$,  
C\'ecile Monthus$^{(2)}$ and  Marc Yor$^{(3)}$  }

\date{ $^{(1)}$  Division de Physique Th\'eorique\thanks{ 
Unit\'e de Recherche des Universit\'es Paris 6 et Paris 11 associ\'ee au CNRS }, 
\ IPN B\^atiment 100, 
 91406 ORSAY C\'edex , FRANCE \\
and L.P.T.P.E., Tour 12, 
 4 Place Jussieu 75252 PARIS Cedex 05, FRANCE \\
\hbox{  }  \hfill \break
$^{(2)}$ Service de Physique Th\'eorique,
C. E.  Saclay, Orme des Merisiers \\
91191 Gif-sur-Yvette C\'edex, FRANCE \\
\hbox{  }  \hfill \break
$^{(3)}$ Laboratoire de Probabilit\'es, Universit\'e Paris 6 \\
 4 Place Jussieu, Tour 56, 75252 PARIS Cedex 05, FRANCE}

\vskip 1 true cm


\maketitle

\begin{abstract}
{ The paper deals with exponential functionals of the linear Brownian motion 
which arise in different contexts such as continuous time  
finance models and one-dimensional disordered models.
We study some properties of these exponential functionals
in relation with the problem of a particle coupled to a heat bath in a
Wiener potential. Explicit expressions for the distribution of the
 free energy are presented.}
\end{abstract}

\vfill

\noindent Electronic addresses : \hfill \break
comtet@ipncls.in2p3.fr  \hfill \break
 monthus@amoco.saclay.cea.fr  

\vskip 2 true cm

\noindent   \hfill November 1995

\newpage

\baselineskip 6mm

Consider a linear Brownian motion $(B_s, s\geq 0, B_0=0)$,
 and a given drift $\mu$. Exponential functionals of the following form
\be
A_t^{(\mu)} = \int_0^t ds \ e^{ \displaystyle - 2(B_s+\mu s)}
\label{atmu}
\ee
have recently been a subject of common interest for mathematicians and for physicists.

Some recent mathematical studies have been partly motivated by continuum-time $t$ 
finance models in which most stock price dynamics are assumed to be driven by the exponential
of a Brownian motion with drift \cite{Duf90} \cite{Gem93}. 
In such studies, the representation  
 \be
e^{ \displaystyle - (B_t+\mu t)} = R \bigg(
\int_0^t ds \ e^{ \displaystyle - 2(B_s+\mu s)} \bigg)
\label{Lamp}
 \ee
where $\big({R(u), u\geq 0}\big)$ is a Bessel process, i.e. an element of an important
class of diffusions, exhibits the importance of the functional $A_t^{(\mu)}$.
Formula (\ref{Lamp}) is a particular instance of the Lamperti relation which
expresses $\big( \exp(\xi_t), t\geq0 \big)$ as
\be
e^{ \displaystyle \xi_t} = X \bigg(
\int_0^t ds \ e^{ \displaystyle 2\xi_s} \bigg)
 \ee
where $\big({X(u), u\geq 0}\big)$ is a semi-stable Markov process (see \cite{Car94}
\cite{Car95} for some applications, partly in mathematical finance).

 In physics, 
these exponential functionals play a central role in the context of 
one dimensional classical diffusion in a random environment. The random variable
$A_{\infty}^{(\mu)}$ can indeed be interpreted as a trapping time. 
Its probability distribution controls the anomalous diffusive 
behaviors of the particle at large time in a infinite sample
\cite{Bou90a} \cite{Bou90b} \cite{Geo88}. The distribution of  
$A_{L}^{(\mu)}$ occurs when studying the maximum reached by
the Brownian particle in a drifted Brownian potential \cite{Kaw93}. 
 The functional $A_{L}^{(\mu)}$ also arises in the study of the transport
 properties of disordered samples of finite length $L$ 
 \cite{Osh93a} \cite{Osh93b} \cite{Mon94}. \hfill \break
 In fact, $A_L^{(\mu)}$ represents the continuous
 space analogue of the random series introduced
 by Kesten {\it et al.} \cite{Kes75} for the so called "random random walk". 
 This random series is generated by a linear recurrence relation with random coefficients 
 and therefore constitutes
 a discrete random multiplicative stochastic process.
 $A_t^{(\mu)}$ also represents a very simple case of a continuous random multiplicative stochastic
 process \cite{Com95} which may be related to hyperbolic Brownian motion \cite{Yor92a}
 \cite{Com95}. 

In this article, we discuss some properties of these functionals, concentrating
ourselves mostly on the mean-value $E \big( \ln A_t^{(\mu)} \big)$, in relation with another
 physical interpretation inspired by the statistical mechanics of disordered systems.
In these systems, the partition function $Z$ is a functional which depends on a set 
of "quenched'' random couplings. In order to obtain the thermodynamic properties 
of the system, one has to compute the average over the disorder of the free-energy $F$
\be
E(F)=E(-kT\ln Z)
\ee
This calculation can rarely be done exactly.
The determination of the probability distribution of $F$ is a still more difficult task.
One of the very few cases for which such a calculation can be done exactly is the 
 Random Energy Model \cite{Der}. It is therefore highly desirable to investigate other 
explicitly solvable cases \cite{Opp93} \cite{Bro95}
where the usual tools of disordered systems such as replica methods and variational 
techniques can be tested \cite{Bro95}.

\section{  Physical motivation : A toy-model for disordered systems}

Let us consider a particle confined on the interval $0 \leq x \leq L$
and submitted to a random force $F(x)$ distributed as a Gaussian white noise around
some mean value $-f_0$. The corresponding random potential is then simply a Brownian motion
 with drift which can be written in terms of the Wiener process $B_x$ as  
\be
U(x)=-\int_0^x F(y) dy =f_0 x +\sqrt{\sigma} B_x
\ee
For a given sample, that is for a given realization of the potential $U(x)$, we
define the partition function
\be
Z_L = \int_0^L dx e^{\displaystyle -\beta U(x)}  
\ee
where as usual $\beta$ is the inverse temperature of the system.
It is convenient to introduce $\alpha = {\beta^2 \sigma \over 2}$ and
the dimensionless parameter $\mu = {2 f_0 \over \beta \sigma}$ to rewrite $Z_L$
as 
\be
Z_L^{(\mu) }=\int_0^L dx \ e^{\displaystyle
 -( \alpha \mu x +\sqrt{2 \alpha} B_x )}
\label{ZLmudef}
\ee
Therefore, for $\alpha=2$,
$Z_L^{(\mu) }$ and $ A_{{ L}}^{(\mu)}$ coincide. For $\alpha \neq 2$,
 using the scaling properties of the Brownian motion, one obtains 
\be
Z_L^{(\mu) } \quad \stackrel{\hbox{\scriptsize(law)}}{=} \quad 
 {2 \over \alpha} \ \int_0^{\displaystyle {\alpha L \over 2}} dx e^{\displaystyle
 - 2( \mu x + B_x )} 
 \ee
 Hence 
 \be
Z_L^{(\mu) } \quad \stackrel{\hbox{\scriptsize(law)}}{=} \quad  
{2 \over \alpha}\  A_{ {\alpha L \over 2}}^{(\mu)}
\label{ZAlaw}
\ee

The thermal average of any function $g(x)$ of the position of the particle 
for a given sample will be denoted by an upper-bar
\be
\overline{g(x)} \equiv \displaystyle { { \int_0^L dx \  g(x) \ e^{\displaystyle -\beta U(x)} } 
\over {\int_0^L dx \ e^{\displaystyle -\beta U(x)}  } }
\ee
For instance the thermal average and variance of the position read  
\be
\overline{x} = {1 \over Z_L^{(\mu) }} \left( -{1 \over \alpha}
{\partial \over \partial \mu} Z_L^{(\mu) } \right)
 =  -{1 \over \alpha} {\partial \over \partial \mu} \ln(Z_L^{(\mu) }) 
\ee
\be
\overline{x^2} -(\overline{x})^2 =
{1 \over \alpha^2} {\partial^2 \over \partial \mu^2} \ln(Z_L^{(\mu) }) 
\ee 
More generally, the generating function of the thermal cumulants of the position reads
\be
\ln \bigg( \overline{ e^{-px}} \bigg) = \ln  Z_L^{(\mu+{p \over \alpha}) } 
- \ln  Z_L^{(\mu) } 
\ee
These relations show that the statistical properties
of the position of the particle in the case $\mu=0$
require in fact the knowledge of
 the partition function with an arbitrary drift $\mu$.

The fundamental quantity for the statistical mechanics of this disordered system 
is  the free energy $F_L^{(\mu) }$ related to the logarithm of the partition function
\be
 F_L^{(\mu) } = -{1 \over \beta} \ln  Z_L^{(\mu) } 
\ee
This paper deals essentially with the statistical properties of the free energy,
and particularly with the mean free energy over the disorder denoted by
\be
E \left( F_L^{(\mu) } \right)= -{1 \over \beta} E \left( \ln  Z_L^{(\mu) } \right)
\ee
This work also gives us the opportunity to bring together various
 results which were scattered both in the physics and mathematics literature.
A comparison between them often yields some interesting identities
with a non trivial probabilistic content.

The paper is organized as follows.
In section \ref{mu0}, we consider  
the case of random potential with zero-drift $\mu=0$. We first give the probability
distribution of the free energy. We then establish
 various formulae for its mean value, using in particular the Bougerol identity, and show 
that the replica method gives the correct result. We also discuss the asymptotic
behavior of the mean free energy in the limit of a large sample $L \to \infty$.
In section \ref{mun0}, we discuss the properties of the free energy in 
the case of a random potential with a positive drift $\mu>0$. We also establish
some relations between $E \left( \ln  Z_L^{(\mu) } \right)$ and 
$E \left( {1 \over { Z_L^{(\nu) }}} \right)$. In section \ref{explength},
we discuss the case where the length of the sample is an independent random variable 
 which is exponentially distributed.

For notations and properties of the special functions appearing in this paper, we
refer the reader to Lebedev \cite{Leb72}.

\vskip 1 true cm

\section{ Case of random potential with zero drift $\mu=0$  }
\label{mu0}

\subsection{ Distribution of the partition function $Z_L^{(0)}$
 and associated free energy } 

The expression of the generating function of $Z_L^{(0) }$
has already been derived in another context 
through the resummation of the series of moments \cite{Osh93a}
and by a path-integral approach \cite{Mon94}
\be
E \bigg( e^{ \displaystyle -p Z_L^{(0) }} \bigg)={2\over\pi} \int_0^\infty dk \ 
\cosh{\pi k\over 2}\ K_{ik} \left(2\sqrt{{p\over\alpha}}\right) 
 \ e^{ \displaystyle -k^2 {\alpha L \over 4} }
\label{gen}
\ee
We  also refer the reader to \cite{Ali95}, where it is shown that this result 
may be derived from the Bougerol formula (see later \ref{Bougerol}).
To invert the Laplace transform (\ref{gen}), it is convenient to start from the integral
representation
\be
K_{ik} \left(2\sqrt{{p\over\alpha}}\right) = \int_0^{\infty} dt \ 
\cos kt \ e^{\displaystyle - 2\sqrt{{p\over\alpha}} \cosh t}
\ee
and to perform the integration over $k$ in (\ref{gen}) to obtain
\be
E \bigg( e^{ \displaystyle -p Z_L^{(0) }} \bigg)
={2\over {\sqrt{\pi \alpha L}} } \ 
 e^{\displaystyle  {\pi^2 \over {4\alpha L} }   } \ 
\int_0^\infty dt \cos \left( { {\pi t} \over {\alpha L} } \right)
 e^{\displaystyle - {t^2 \over {\alpha L}}   }
\  e^{\displaystyle - 2\sqrt{p \over \alpha} \cosh t}
\label{genbis}
\ee
We may then use the following identity
\be
e^{\displaystyle - 2\sqrt{{p\over\alpha}} \cosh t} =
{ {\cosh t } \over \sqrt{\pi \alpha}  } \int_0^{\infty} \ 
 {dZ \over Z^{3 \over 2}} \ e^{\displaystyle -pZ}
\ e^{\displaystyle  -{ {\cosh^2 t } \over { \alpha Z}  } }
\ee
to cast (\ref{genbis}) into the form
\be
E \bigg( e^{ \displaystyle -p Z_L^{(0) }} \bigg) =\int_0^{\infty} dZ \ e^{-pZ} \ 
\psi_L^{(0) } (Z)
\ee
where 
\be
\psi_L^{(0) } (Z) = { {2 e^{ { \pi^2 \over{4\alpha L}}   }}
 \over {\pi \alpha \sqrt{ L}}} \ 
 {1 \over Z^{3 \over 2}} \ 
  \int_0^\infty dt 
\cosh t \ \cos \left({{\pi t} \over {\alpha L} } \right)
 e^{\displaystyle - {t^2 \over{\alpha L} }   }  \ 
 e^{\displaystyle  -{ {\cosh^2 t } \over { \alpha Z}  } }
\label{psiz}
\ee
denotes the probability distribution of the partition function $ Z_L^{(0) }$.
A simple change of variables then gives the probability distribution
$P_L^{(0) } (F)$ of the free energy 
 $F_L^{(0) } = -{1 \over \beta} \ln  Z_L^{(0) }$ 
\be
P_L^{(0) } (F) = { { 2\beta e^{ { \pi^2 \over{4\alpha L}}   }}
 \over {\pi \alpha \sqrt{ L}}} \ 
 e^{\displaystyle  {\beta \over 2}F } \ 
  \int_{0}^{\infty} dt 
\cosh t \ \cos \left({{\pi t} \over {\alpha L} } \right)
 e^{\displaystyle - {t^2 \over{\alpha L} }   }  \ 
 e^{\displaystyle  - \left({ {\cosh^2 t } \over \alpha   } \right) e^{\beta F} }
\label{Pf}
\ee
It is interesting to compare with a similar formula given in \cite{Bro95}.

In the limit of large $L$, the probability density $X_L(\xi)$ of the reduced variable 
$\xi={ {\beta F - \ln \alpha} \over { \sqrt{2 \alpha L}}}$ tends to the Gaussian 
\be
X_{L } (\xi) \operarrow_{L \to \infty} {1 \over \sqrt{ 2 \pi}}
  e^{\displaystyle  - { \xi^2 \over 2} } 
\ee
This asymptotic result may in fact be directly obtained
 since from the definition (\ref{ZLmudef})
\be
Z_L^{(0) }= \int_0^L dx \ e^{\displaystyle
  +\sqrt{2 \alpha} B_x }
 \quad \stackrel{\hbox{\scriptsize(law)}}{=} \quad 
 \  L \int_0^{ 1} ds \ e^{\displaystyle \sqrt{2 \alpha L} B_s} 
 \ee
hence
\be
{1 \over \sqrt{L}} \ln Z_L^{(0) } = {\ln L \over \sqrt{L}}+ \ln \bigg(
\int_0^{ 1} ds \  e^{\displaystyle \sqrt{2 \alpha L} B_s}  \bigg)^{{1 \over \sqrt{L}}} 
\ee
Since
\be
\ln \bigg(
\int_0^{ 1} ds e^{\displaystyle \sqrt{2 \alpha L} B_s}  \bigg)^{{1 \over \sqrt{L}}} 
 \operarrow_{L \to \infty}\ln \  e^{\displaystyle \sqrt{2 \alpha } \sup_{s \leq 1} B_s}
 \quad \stackrel{\hbox{\scriptsize(law)}}{=} \quad  \sqrt{2 \alpha } \vert N \vert
\ee
where $N$ is a normalized Gaussian variable, one has
\be 
{1 \over \sqrt{2 \alpha L}} \ln Z_L^{(0) } \quad \stackrel{\hbox{\scriptsize(law)}}
{\operarrow_{L \to \infty} } \quad \vert N \vert
\ee

\subsection{ An expression of $E \bigg(\ln  Z_L^{(0) } \bigg)$ derived 
from the generating function } 

The Frullani identity 
\be
\ln  Z_L^{(0)}= \int_0^{\infty} {dp \over p} \ 
\bigg[ e^{ \displaystyle -p} - e^{\displaystyle - p Z_L^{(0)}} \bigg]  
\ee
may be used to compute the mean of the logarithm of $ Z_L^{(0) }$
from the generating function of equation (\ref{gen}). 
Using the intermediate regularization
\be
E \bigg( \ln \big( Z_L^{(0) } \big) \bigg) = \lim_{\epsilon \to 0^+}
\bigg[ \Gamma(\epsilon) - \int_0^{\infty} {dp } \ p^{\displaystyle \epsilon -1} \ 
E \bigg(  e^{\displaystyle - p Z_L^{(0)}} \bigg) \bigg]
\label{reg}
\ee
one obtains
\be
E \bigg( \ln  Z_L^{(0) }  \bigg) = {2 \over \pi}
\int_0^{\infty} {dk \over k^2} \ \bigg[1- e^{\displaystyle - \alpha L k^2}
\pi k \coth(\pi k) \bigg]+C-\ln \alpha
\label{lnZcoth}
\ee
where $C=-\Gamma'(1)$ is Euler's constant.

\vskip 0.5 true cm

\subsection{  Bougerol's identity}

\vskip 0.5 true cm

Bougerol's identity \cite{Bou83} is an identity in law relating two independent Brownian motions
$(B_s, s\geq 0)$ and $(\gamma_u, u\geq 0)$, and involving the exponential functional
we are interested in. The statement is that for fixed $t$
\be
\sinh (B_t) \quad \stackrel{\hbox{\scriptsize(law)}}{=} \quad \gamma_{ A_t } \hskip 1 true cm \hbox{where} \hskip 1 true cm
A_t = \int_0^t ds \  e^{ \displaystyle - 2 B_s} =A_t^{(0)}  
\label{Bougerol}
\ee
In the Appendix, we present a simple proof of this identity due to L. Alili,
 and we refer the reader to \cite{Ali95} for further details and
possible generalizations for the case of a non-vanishing drift $\mu \neq 0$.

Scaling properties of Brownian motion then give
\be
A_t \quad \stackrel{\hbox{\scriptsize(law)}}{=} \quad 
 t \int_0^1 du \  e^{\displaystyle - 2 \sqrt{t} B_u} 
\ee 
\ba
&\sinh (B_t)  \quad \stackrel{\hbox{\scriptsize(law)}}{=} \quad
\ \left({ \int_0^1 du \  e^{\displaystyle - 2 \sqrt{t} B_u}  }\right)^{1 \over 2}
  \gamma_t  \\  
&  \stackrel{\hbox{\scriptsize(law)}}{=} \quad \  \left({ A_t \over t}\right)^{1 \over 2} \ \gamma_t
\label{BougeConseq}
\ea
It follows that
\be
E\bigg( \ln A_t  \bigg) =
E \bigg( \ln \left({ \sinh (B_t) \over B_t} \right)^2 \bigg) +\ln t 
= \int_{-\infty}^{+\infty} {dx \over \sqrt{2\pi t}}  \ e^{\displaystyle - {x^2 \over 2t}} \
\ln \left({ \sinh (x) \over x} \right)^2 +\ln t  
\ee
Starting from the partition function 
\be
Z_L^{(0)} \quad \stackrel{\hbox{\scriptsize(law)}}{=} \quad
 {2 \over \alpha} \ A_{ \displaystyle{ {\alpha L} \over 2} }
\ee
we thus get a new expression for the mean value of the logarithm
\be
E \bigg( \ln  Z_L^{(0) } \bigg) = 
 \int_{-\infty}^{+\infty} {dx \over \sqrt {\pi \alpha L} } \ e^{\displaystyle
- { x^2 \over {\alpha L}} } \
\ln \left({ \sinh (x) \over x} \right)^2 + \ln (L)
\label{lnZBou}
\ee

Comparison with our previous result (\ref{lnZcoth}) leads to the identity
\be
2 \int_{-\infty}^{+\infty} {dx \over \sqrt{\pi \alpha L} } \ e^{\displaystyle
- { x^2 \over {\alpha L}} } \
\ln \left({ \sinh (x) \over x} \right)  = 
{2 \over \pi}
\int_{-\infty}^{+\infty} {dk \over k^2} \ \bigg[1- e^{- \alpha L k^2}
\pi k \coth(\pi k) \bigg] +C - { \ln (\alpha L) } 
\ee
In order to understand the meaning of this identity, we differentiate
both sides with respect to $L$
and use the heat equation
\be
{\partial \over \partial L} \bigg( {1\over \sqrt{\pi \alpha L} } \ 
e^{\displaystyle -{x^2 \over {\alpha L}}}
\bigg) = {\alpha \over 4} {\partial^2 \over \partial^2 x}
\bigg( {1\over \sqrt{\pi \alpha L} } \ e^{\displaystyle - {x^2 \over {\alpha L}}} \bigg)
\ee
After an easy integration by part we obtain 
\be
\int_{-\infty}^{+\infty} {dx \over \sqrt {\pi \alpha L} } \ e^{\displaystyle - {
x^2 \over {\alpha L}}} \
\bigg[ {\partial^2 \over \partial^2 x}  \ln \left({ \sinh (x) \over x} \right) \bigg]  =
   {\alpha } \int_{-\infty}^{+\infty} {dk } \  e^{\displaystyle - \alpha L k^2}
\bigg[  k \coth(\pi k)  - \vert k \vert \bigg] 
\ee
This identity is a particular instance of Plancherel's formula, since the Fourier
transform of the function
\be
{\partial^2 \over \partial^2 x}  \ln \left({ \sinh (x) \over x} \right)
= {1 \over  x^2} -{1 \over \sinh^2 (x)}
\ee
can be easily obtained after an integration in the complex plane
\be
\int_{-\infty}^{+\infty} dx \ e^{i k x}
\bigg[ {1 \over  x^2} -{1 \over \sinh^2 (x)} \bigg] = \pi 
\bigg[  k \coth \left({\pi \over 2}  k\right)  - \vert k \vert \bigg]
\label{TFpc}
\ee
This formula is also encountered as follows in the study \cite{Bia87}
of the Hilbert transform of Brownian motion
\be
H_u = \lim_{\epsilon \to 0^+} \int_0^{u} {ds \over B_s} 
1_{\left(\vert B_s \vert \geq \epsilon \right)}
\ee
If n$(d\epsilon)$ denotes the characteristic measure of Brownian excursions,
and $\epsilon$ the generic excursion with lifetime
  $V(\epsilon) = \inf\{t >0 : \epsilon(t)=0\}$, then the law of  $H_V 
\equiv \int_0^{V} {ds \over \epsilon_s}$ under n is ${\pi dx \over x^2}$, 
and formula (\ref{TFpc}) may be interpreted as
\be
\pi \bigg( \lambda \coth \left( {\pi \lambda \over \theta} \right) -\vert \lambda \vert
\bigg) =
\pi \int_{-\infty}^{+\infty} {dx \over x^2} \ e^{i \lambda x} 
\ \hbox{n} \bigg( 1-e^{-\theta^2 V \over 2} \bigg\vert H_V=x \bigg) 
\ee
with
\be
\hbox{ n} \bigg( 1-e^{-\theta^2 V \over 2} \bigg\vert H_V=x \bigg) 
= 1- \left(
{   {\theta x \over 2} \over {\sinh \left( {\theta x \over 2} \right) }}
\right)^2
\ee
in the particular case $\theta=2$.

\vskip 0.5 true cm

\subsection{ Mean free energy through Replica method }

Let $X=Z_L^{(0) }$. The replica method is based on the identity
\be
E \big[ \ln X   \big] = \lim_{n \to 0}
{ { E \big[ X^n \big] -1} \over n}
\ee
In many applications, one proceeds by looking for
an analytic continuation
$n \to 0$ of the expressions of integer moments of $X$.  
This procedure is in general mathematically ill-founded, 
but in the present case non-integer moments can be computed. 

The integer moments of the partition function are well-known
\cite{Duf90} \cite{Yor92a} \cite{Osh93a} \cite{Mon94}
\be
E \bigg[ \big (Z_L^{(0) } \big)^n \bigg] =
{1 \over \alpha^n } \ \sum_{k=1}^n \ e^{\alpha
L k^2}(-1)^{n-k}{\Gamma(n)\over\Gamma(2n)}C_{n}^k+{(-1)^n\over n!}
\ee
Since this expression contains a sum of $n$ terms, it has a priori
no meaning for non-integer $n$. 
However, the following integral representation
for the moment of order $n$ \cite{Osh93a}
 \be
 E \bigg[ \big (Z_L^{(0) } \big)^n \bigg] = {2 }
{\Gamma({1 \over 2}) \over \Gamma(n+{1 \over 2})}
\int_0^{\infty} {dx \over {\sqrt{\pi \alpha L}}} \ e^{ \displaystyle 
- {x^2 \over {\alpha L}}} \ \left( { \sinh^2 x \over \alpha}  \right)^{n} 
\ee
is valid for any positive real $n \geq 0$, as can be shown using
the consequence (\ref{BougeConseq}) of Bougerol's identity (\ref{Bougerol}).

The following expansions in $n$, as $n \to 0$ 
\be
\left( { \sinh^2 x \over \alpha}  \right)^{n} =
1+n\ln \left( { \sinh^2 x \over \alpha}  \right)+o(n)
\ee
\be
{\Gamma({1 \over 2}) \over \Gamma(n+{1 \over 2})} = 1-n {\Gamma'({1 \over 2})
\over \Gamma({1 \over 2}) } +o(n) =1+n \big(C+2\ln2 \big)+o(n)
\ee
where $C=-\Gamma'(1)$ denotes Euler constant, lead to the expression
\be
E \bigg[ \ln \big( Z_L^{(0) } \big) \bigg] = \lim_{n \to 0}
{ { E \bigg( \big (Z_L^{(0) } \big)^n \bigg) -1} \over n}
\ee
\be
= C+2\ln2 - \ln \alpha  + \int_{-\infty}^{+\infty} 
{dx \over {\sqrt{\pi \alpha L}}} \ e^{\displaystyle - {x^2 \over {\alpha L}}}
\ \ln \big( { \sinh^2 x }  \big)
\label{lnZrep}
\ee
This formula is of course directly related to the previous expression (\ref{lnZBou}) 
obtained using Bougerol's identity since 
\be
-  \int_{-\infty}^{+\infty} {dx \over \sqrt {\pi \alpha L}} 
\ e^{\displaystyle - {x^2 \over {\alpha L}}} \
\ln (x^2) + \ln (L) = C+2\ln2 - \ln \alpha 
\ee

\vskip 0.5 true cm

\subsection{ Asymptotic expression of the mean free energy for large $L$  }

The various expressions obtained previously (\ref{lnZcoth}-\ref{lnZBou}-\ref{lnZrep})
yield the asymptotic behavior
\be
E \bigg[ \ln \big( Z_L^{(0) } \big) \bigg]=
2 \sqrt{ {\alpha L} \over \pi} + C - \ln \alpha  -
{ \pi^{3 \over 2} \over { 3 \sqrt{\pi \alpha L}}}
 +O \left( { 1 \over { ( \alpha L)^{3 \over 2} }} \right)
\qquad \hbox{as} \ \ L \to \infty 
\label{meanasy}
\ee
which agrees with \cite{Bro95}.

We now show how the first two terms may in fact be recovered through a direct asymptotic analysis
 of
\be
Z_L^{(0) }=\int_0^L dx \ e^{\displaystyle
 \sqrt{2 \alpha} B_x } 
\ee
Using the scaling properties of Brownian motion, one has
\be
Z_L^{(0) } \quad \stackrel{\hbox{\scriptsize(law)}}{=} \quad 
 {1 \over {2 \alpha} }\int_0^{2 \alpha L} dx \ e^{\displaystyle  B_x } 
\ee
It is convenient to set $\Lambda=2 \alpha L$ and to define
\be
I_{\Lambda}= E \bigg[ \ln \int_0^{\Lambda} dx \ e^{\displaystyle  B_x } \bigg]
\ee
Introducing the one-sided supremum 
$S_{\Lambda} = \displaystyle \sup_{x \leq \Lambda} B_x$, one has
\be
I_{\Lambda}=E \big( S_{\Lambda} \big)+
E \bigg[ \ln  \int_0^{\Lambda} dx \ e^{\displaystyle
-(S_{\Lambda}- B_x)   } \bigg] 
\ee
It follows from the scaling properties of Brownian motion that 
\be
I_{\Lambda}= \sqrt{\Lambda} E \big( S_1 \big)+
E \bigg[ \ln  \big( {\Lambda} \int_0^1 dx \ e^{\displaystyle
 - \sqrt{\Lambda}  (S_1- B_x) } \big)   \bigg]
\ee
Recall that from the reflection principle $S_1 
 \quad \stackrel{\hbox{\scriptsize(law)}}{{=}} \quad
\vert B_1 \vert$. This implies 
\be
E \big( S_1 \big) =E \big( \vert B_1 \vert \big) = \sqrt{2 \over \pi}
\ee
For the remaining part of $I_{\Lambda} $, theorem $1$ of Pitman and Yor \cite{Pit93} gives 
the convergence in law
\be
 {\Lambda} \int_0^1 dx \ e^{\displaystyle
 -  \sqrt{\Lambda} (S_1- B_x)  } 
 \quad \stackrel{\hbox{\scriptsize(law)}}{\operarrow_{\Lambda \to \infty}} \quad 
 4 \big( T_1+ \hat{T}_1  \big)   
\ee
where $T_1$ and $\hat{T}_1$ are two independent copies of the 
first hitting time of $1$ by a two-dimensional Bessel process starting from $0$.
Hence
\be
E \bigg[ \ln  \big( {\Lambda} \int_0^1 dx \ e^{\displaystyle
 - \sqrt{\Lambda} (S_1- B_x) } \big)   \bigg] \operarrow_{\Lambda \to \infty}
E \bigg[ \ln 4 \big( T_1+ \hat{T}_1  \big)  \bigg] 
\label{PY93}
\ee
The right-hand side of last equation
can be evaluated either by direct calculation or by appealing to recent results related to the 
"agreement formula" for Bessel processes \cite{Pit95}. The direct calculation 
proceeds as follows : one starts
 again with the identity 
\be
E \bigg[ \ln  \big( T_1+ \hat{T}_1  \big)   \bigg]
= \int_0^{\infty} {du \over u} 
\bigg[ e^{\displaystyle -u} - 
E \left(e^{-\displaystyle u  \big( T_1+ \hat{T}_1  \big) } \right)  \bigg]
\ee
and uses the expression of the Laplace transform \cite{Ken78}
\be
E \left(e^{-\displaystyle u   T_1  } \right) = {1 \over {I_0(\sqrt{2u})}}  
\ee
After an intermediate regularization, one obtains
\ba
E \bigg[ \ln  \big( T_1+ \hat{T}_1  \big) \bigg]
& \displaystyle = \lim_{\epsilon \to 0^+} \int_{\epsilon}^{\infty} {du \over u} 
\bigg[ e^{\displaystyle -u} -  {1 \over {I_0^2(\sqrt{2u})}} \bigg] \\
& \quad \displaystyle = \lim_{\epsilon \to 0^+} \bigg[ -C- \ln {\epsilon^2 \over 2}
-2{K_0(\epsilon )\over I_0(\epsilon )} \bigg] = C-\ln 2
\label{lnT1T1}
\ea
The other method relies on the following consequence of 
the "agreement formula" for Bessel processes ( equation (37) of \cite{Pit95})
\be
E_{\delta} \bigg[  \big( T_1+ \hat{T}_1  \big)^{{ \delta \over 2} -1} \bigg] =
{1 \over { 2^{\mu} \Gamma(1+\mu) } }  
\label{agree}
\ee
Here $\delta =2(1+\mu)$ is the dimension of the Bessel process, and $E_{\delta}$
denotes the expectation with respect to the law of this process, starting at $0$. ( For $\delta=2$, or equivalently $\mu=0$, we simply note $E$ instead of $E_2$).
By differentiating both sides of (\ref{agree}) with respect to $\mu$ at $\mu=0$,
one recovers (\ref{lnT1T1})
\be
E \bigg[ \ln  \big( T_1+ \hat{T}_1  \big)   \bigg] = C-\ln 2
\label{lnT12}
\ee
Finally, combining (\ref{PY93}) and (\ref{lnT12}), we obtain  
\be
I_{\Lambda}- \sqrt{2{\Lambda} \over \pi}\operarrow_{\Lambda \to \infty}
 C+\ln 2 
\ee
We therefore recover the first two terms of (\ref{meanasy})
\be
E \bigg[ \ln \big( Z_L^{(0) } \big) \bigg]- \bigg(
2 \sqrt{ {\alpha L} \over \pi} + C - \ln \alpha \bigg)
 \operarrow_{L \to \infty} 0
\ee

\vskip 1 true cm

\section{  Case of random potential with drift $\mu > 0 $  }
\label{mun0}

\vskip 0.5 true cm
\subsection{  Distribution of the partition function $Z_L^{(\mu) } $ 
and associated free energy }

The probability distribution $\psi_L^{(\mu)}(Z)$ of the partition function
$Z_L^{(\mu) } $ has already been obtained in another context as the solution
of the Fokker-Planck equation \cite{Mon94} \cite{Com95}
\be
{\partial \psi_L^{(\mu)}(Z) \over \partial L} = {\partial \over \partial Z}
\bigg[ \alpha Z^2 {\partial \psi_L^{(\mu)}(Z) \over \partial Z} +
 \bigg( (\mu + 1)\alpha Z - 1 \bigg) \psi_L^{(\mu)}(Z) \bigg] 
\label{FPpsi}
\ee
satisfying the initial condition $\psi_{L=0}^{(\mu)}(Z) =\delta(Z)$.
The eigenfunction expansion of the solution 
 exhibits the following relaxation spectrum with the length $L$ 
\ba
&\psi_L^{(\mu)}(Z)= \displaystyle \alpha \sum_{0\leq n<{\mu\over 2}}
e^{ \displaystyle -\alpha Ln(\mu-n)}
{(-1)^n(\mu-2n)\over \Gamma(1+\mu-n)}
\left(
{1\over \alpha Z}
\right)^{1+\mu-n}
L_n^{\mu-2n}
\left(
{1\over \alpha Z}
\right)e^{- \displaystyle {1 \over \alpha Z}}
\label{psiZmu} \\
&\displaystyle +{\alpha\over 4\pi^2}\int_0^{\infty} \! ds\ 
e^{ \displaystyle -{\alpha L\over 4}(\mu^2+s^2)}
s\sinh \pi s
\left\vert
\Gamma\left(-{\mu\over 2}+i {s\over 2}\right)
\right\vert^2
\left({1\over \alpha Z}\right)^{1+\mu\over 2} \!
W_{{1+\mu\over 2},i{s\over 2}}
\left({1\over \alpha Z}\right)e^{-\displaystyle {1\over 2\alpha Z}}\ 
\label{psiZmubis}
\ea
where $L_n^\alpha$ are Laguerre's polynomials and $W_{p,\nu}$ are Whittaker's
functions.
Contrary to the case $\mu=0$, there exists for $\mu>0$
 a limit distribution as $L \to \infty$ 
\be
\psi_{\infty}^{(\mu)}(Z)= 
{\alpha  \over \Gamma(\mu)}
\left( {1\over \alpha Z} \right)^{1+\mu}
e^{- \displaystyle {1 \over \alpha Z}} \\
\label{trappingtime}
\ee
In the physics literature, this limit distribution
 was first obtained in the context
of one dimensional Brownian diffusion in a Brownian drifted potential,
where $Z_{\infty}^{(\mu)}$ plays the role of an effective trapping time 
\cite{Bou90b}, and was recently rediscovered in another context \cite{Opp93}.
In the mathematics literature, this limit distribution
\be
A_{\infty}^{(\mu) } \quad \stackrel{\hbox{\scriptsize(law)}}{=} \quad 
{1 \over {2Y_{\mu}}}
\ee
where $Y_{\mu}$ is a Gamma variable with parameter $\mu$
\be
P \bigg( Y_{\mu} \in (y,y+dy) \bigg) = {dy \over \Gamma(\mu)} \ y^{\mu-1} \ e^{-y}
\label{gamma}
\ee
 was first obtained 
by Dufresne \cite{Duf90}, then shown in \cite{Yor92c} to be another expression
of the law of last passage times for transient Bessel processes. 
It is interesting to point out that equation (\ref{FPpsi}) first appeared 
in a paper of Wong 
\cite{Won64} (see also \cite{Hon86}). However, the precise connection with our work is
beyond the scope of this paper.

\subsection{  Mean free energy $E \bigg( \ln Z_L^{(\mu) }  \bigg) $ }

The Laplace transform of $Z_L^{(\mu)}$ corresponding 
to the probability distribution (\ref{psiZmu}-\ref{psiZmubis}) reads \cite{Mon94}
\ba
&E \bigg( e^{-p Z_L^{(\mu) }}  \bigg) = \displaystyle \sum_{0\leq n<{\mu\over 2}}
e^{ \displaystyle -\alpha Ln(\mu-n)}
{2(\mu-2n)\over n!\Gamma(1+\mu-n)}
\left(
{p\over \alpha}
\right)^{\mu/2} K_{\mu-2n}
\left(
2\sqrt{p\over\alpha}\right) \\
&+\displaystyle {1\over 2\pi^2}\int_0^{\infty}ds\ e^{\displaystyle -{\alpha L\over 4}(\mu^2+s^2)}
s\sinh \pi s
\left\vert
\Gamma\left(-{\mu\over 2}+i {s\over 2}\right)
\right\vert^2
\left(
{p\over \alpha}
\right)^{\mu/2} K_{is}\left(
2 \sqrt{p\over\alpha}\right)\ 
\label{genmu}
\ea
We use again the Frullani identity
\be
E\bigg( \ln  Z_L^{(\mu) } \bigg) = \int_0^{\infty} {dp \over p} \ 
\bigg[ e^{ \displaystyle -p} - E\left( e^{\displaystyle - p Z_L^{(\mu)}} \right) \bigg]  
\ee
to obtain after an intermediate regularisation as in (\ref{reg})
\ba
&E\bigg( \ln  Z_L^{(\mu) } \bigg) = \displaystyle
-\ln \alpha -{ {\Gamma'(\mu)} \over{\Gamma(\mu)} }
- \sum_{1\leq n <{\mu\over 2}} { {\mu-2n} \over { n(\mu-n)}} \ e^{ \displaystyle 
-\alpha L n(\mu-n)}   \\
&\displaystyle -2 \int_0^{\infty} ds\ 
{s\over {\mu^2+s^2}} \ { \sinh\pi s \over {\cosh \pi s -\cos\pi\mu}}\ 
e^{ \displaystyle -{\alpha L\over 4}(\mu^2+s^2)} 
\ea
\label{ElnmuLexp}
Obviously, the two first terms correspond to the contribution of
the limit distribution (\ref{trappingtime})
\be
E\bigg( \ln  Z_{\infty}^{(\mu) } \bigg) = \displaystyle
-\ln \alpha -{ {\Gamma'(\mu)} \over{\Gamma(\mu)} }
\label{Elnmuinf}
\ee
Now, it is easily deduced from (\ref{ElnmuLexp}) and (\ref{Elnmuinf}) that
\be
E\bigg( \ln  Z_{\infty}^{(\mu) } \bigg) - E\bigg( \ln  Z_L^{(\mu) } \bigg)
\opsimeq_{L \to \infty} 
\left\{ \begin{array}{ll} 
\displaystyle  \left({ {\mu-2} \over {\mu-1}} \right) \ e^{\displaystyle - \alpha (\mu-1) L}
 \quad & \mbox{if $\mu>2$}  \\
\displaystyle  {1 \over \sqrt{ \pi \alpha L}} \  e^{\displaystyle - \alpha  L}  
\quad & \mbox{if $\mu=2$} \\
\displaystyle {  1 \over {\mu^2 (1-\cos \pi \mu)}} 
\left({\pi \over {\alpha L}} \right)^{3 \over 2}   \ 
e^{\displaystyle - { \alpha \mu^2 \over 4}  L}  \quad & \mbox{if $\mu<2$} 
\end{array} \right.
\label{Elnmu1}
\ee 
It is interesting to compare these results to 
those obtained in \cite{Kaw93} for the limiting distribution of the maximum of a 
diffusion process in a Brownian drifted environment. The latter 
  may be stated as
\be
E\bigg( {  {A_{\infty}^{(\mu)} } 
\over {A_{\infty}^{(\mu)}+\tilde{A}_L^{(-\mu)}Ê}} \bigg) 
\opsimeq_{L \to \infty} 
\left\{ \begin{array}{ll} 
\displaystyle \left({ {\mu-2} \over {\mu-1}} \right) \ e^{\displaystyle - 2 (\mu-1) L}
 \quad & \mbox{if $\mu>2$}  \\
\displaystyle {1 \over \sqrt{2 \pi L}} \  e^{\displaystyle - 2  L}  
\quad & \mbox{if $\mu=2$} \\
\displaystyle {  K \over {(2 L)^{3 \over 2} }}  \ 
e^{\displaystyle - {\mu^2 \over 2}  L}  \quad & \mbox{if $\mu<2$} 
\end{array} \right.
\label{KT}
\ee 
where $A_{\infty}^{(\mu)}$ and $\tilde{A}_L^{(-\mu)}$ are two independant
 functionals of type $A_t^{(\nu)}$ defined in (\ref{atmu}). 
The constant $K$ given in \cite{Kaw93} 
\footnote{In fact (\ref{cteKW}) corrects 
the formula for the constant $K$ found in \cite{Kaw93}. The need to divide by 
$\sqrt \pi$ is due to a misprint in \cite{Yor92a}, where on page 528,
 just after formula (6.e), one should read  ${1 \over {(2 \pi^3 t)^{1/2}}} \cdots$
instead of ${1 \over {(2 \pi^2 t)^{1/2}}} \cdots$} 
as the following 4-fold integral 
\be
K={2^{ 1-\mu} \over {\sqrt \pi}} {1 \over \Gamma(\mu)} \int_0^{\infty} dy \  y^{\mu}  
\int_0^{\infty} dz \ e^{- {z \over 2} ( 1+y^2)}\int_0^{\infty} da \ a^{\mu-1}
 {z \over {z+a}} \ e^{- {a \over 2}} \int_0^{\infty} du \ u \sinh u \ e^{- zy \cosh u}  
\label{cteKW}
\ee
may in fact be explicitely computed to give in the end the simple result
\be
K= \pi^{3 \over 2}  { {\Gamma({\mu \over 2})}^2 \over {\Gamma(\mu)}}
{1 \over {(1-\cos \pi \mu)}}
\ee
 For $\mu \geq 2$, the result (\ref{KT}) therefore coincide
with the result (\ref{Elnmu1}), where
 for the particular value $\alpha=2$,
 $Z_L^{(\mu) }$ reduces to $A_L^{(\mu) }$ 
(see (\ref{ZLmudef}) and (\ref{atmu})).
To understand this coincidence, we write
\be
 A_{\infty}^{(\mu) }  =  A_L^{(\mu) } + 
e^{\displaystyle-2(B_L+\mu L)} \  \tilde{A}_{\infty}^{(\mu) }
\ee
where $\tilde{A}_{\infty}^{(\mu) }$ is a variable distributed
 as $A_{\infty}^{(\mu) }$ and independent of $ A_L^{(\mu) }$.
Therefore
\be
E\bigg( \ln  A_{\infty}^{(\mu) } \bigg) - E\bigg( \ln  A_L^{(\mu) } \bigg)
=E\bigg( \ln \left[ 1+ e^{\displaystyle-2(B_L+\mu L)}
  { {\tilde{A}_{\infty}^{(\mu) }} \over {ÊA_L^{(\mu) } }} \right] \bigg)
=E\bigg( \ln \left[ 1+ 
  { {\tilde{A}_{\infty}^{(\mu) }} \over {ÊA_L^{(-\mu) } }} \right] \bigg)
\ee
The comparison between (\ref{Elnmu1}) and (\ref{KT}) 
therefore leads to
\be
E\bigg( \ln \left[ 1+ 
  { {\tilde{A}_{\infty}^{(\mu) }} \over {ÊA_L^{(-\mu) } }} \right] \bigg)
\opsimeq_{L \to \infty}
E\bigg(  { {\tilde{A}_{\infty}^{(\mu) }} \over {
\tilde{A}_{\infty}^{(\mu) } +ÊA_L^{(-\mu) } }} \bigg)  \qquad \hbox{for} \quad \mu \geq 2
\label{mubig2}
 \ee
This is likely to be understood as a consequence of the following plausible statement  
\be
\hbox{If} \qquad  X_n \quad \stackrel{\hbox{\scriptsize(a.s.)}}{\operarrow}  \quad 0,
\quad \hbox{then}
 \quad E \bigg[\ln (1+X_n) \bigg] \sim E \bigg[ {X_n \over {1+X_n}Ê} \bigg]
\label{lem}
\ee
which presumably holds for a large class of random variables $\{X_n\}$, but the precise 
conditions of validity of (\ref{lem}) elude us. 
However, (\ref{mubig2}) does not hold for $\mu <2$ since, in this case, the prefactors 
in (\ref{Elnmu1}) and (\ref{KT}) differ.

To go further into the comparison, we have computed exactly the quantity
occuring in (\ref{KT}) for arbitrary $L$. We start from the identity
\be
E\bigg( {  {Z_{\infty}^{(\mu)} } 
\over {  Z_{\infty}^{(\mu)}+\tilde{Z}_L^{(-\mu)}Ê}  } \bigg) 
=\int_0^{\infty} dp \ E \bigg( Z_{\infty}^{(\mu)} 
 \ e^{\displaystyle - p Z_{\infty}^{(\mu)}} \bigg)
\ \ E\bigg(   e^{\displaystyle - p \tilde{Z}_{L}^{(-\mu)} }\bigg)
\ee
Using the following consequences of (\ref{genmu})
\be
E \bigg( Z_{\infty}^{(\mu)} e^{-p Z_{\infty}^{(\mu) }}  \bigg) =
{2 \over {\alpha \Gamma(\mu)}}
\left(
{p\over \alpha}
\right)^{{\mu-1} \over 2} K_{\mu-1}
\left(
2\sqrt{p\over\alpha}\right) \\
\ee
and
\be
E \bigg( e^{-p \tilde{Z}_L^{(-\mu)}}  \bigg) =  {1\over 2\pi^2}
\int_0^{\infty}ds\ e^{\displaystyle -{\alpha L\over 4}(\mu^2+s^2)}
s\sinh \pi s
\left\vert
\Gamma\left({\mu\over 2}+i {s\over 2}\right)
\right\vert^2
\left(
{p\over \alpha}
\right)^{-\mu/2} K_{is}\left(
2 \sqrt{p\over\alpha}\right)\ 
\label{gen-mu}
\ee
we get 
\be
E\bigg( {  {Z_{\infty}^{(\mu)} } 
\over {Z_{\infty}^{(\mu)}+\tilde{Z}_L^{(-\mu)}Ê}} \bigg) 
=  {1\over {\pi^2 \Gamma(\mu)}} \int_0^{\infty} dx K_{\mu-1}(x)
\int_0^{\infty}ds\ e^{\displaystyle -{\alpha L\over 4}(\mu^2+s^2)}
s\sinh \pi s
\left\vert
\Gamma\left({\mu\over 2}+i {s\over 2}\right)
\right\vert^2
 K_{is} (x) 
\label{maxlmu}
\ee
For $\mu \leq 2$, the order of integrations may be exchanged to give
\be
E\bigg( {  {Z_{\infty}^{(\mu)} } 
\over {Z_{\infty}^{(\mu)}+\tilde{Z}_L^{(-\mu)}Ê}} \bigg) 
=  {1\over {2  \Gamma(\mu)}} 
\int_0^{\infty}ds\ e^{\displaystyle -{\alpha L\over 4}(\mu^2+s^2)}
{{Ês\sinh \pi s} \over {\cosh \pi s - \cos \pi \mu}}
\left\vert
\Gamma\left({\mu\over 2}+i {s\over 2}\right)
\right\vert^2
\label{maxmu-2}
\ee
which reduces for $\mu=2$ to
\be
E\bigg( {  {Z_{\infty}^{(2)} } 
\over {Z_{\infty}^{(2)}+\tilde{Z}_L^{(-2)}Ê}} \bigg) 
=  {1\over 2} 
\int_0^{\infty} ds\ e^{\displaystyle -{\alpha L\over 4}(4+s^2)} \ 
{{Ê\pi s^2 \cosh{ \pi s \over 2} } \over { \sinh^2{ \pi s \over 2}  }}
\label{maxmu=2}
\ee 
For $\mu>2$, one cannot exchange the order of integrations 
in (\ref{maxlmu}). However,
one may start from a series representation ( eq (5.9) in \cite{Mon94})
of the generating function (\ref{gen-mu})
  to obtain, after some algebra involving
deformation of a contour integral in the complex plane
(see \cite{Mon94} for a similar approach) 
the general result for 
arbitrary $\mu >0$ 
\ba
&E\bigg( {  {Z_{\infty}^{(\mu)} } 
\over {Z_{\infty}^{(\mu)}+\tilde{Z}_L^{(-\mu)}Ê}} \bigg) 
 = \displaystyle
 \sum_{1\leq n <{\mu\over 2}} (\mu-2n) 
{ {\Gamma(n) \Gamma(\mu-n)} \over {\Gamma(\mu)}} \ e^{ \displaystyle 
-\alpha L n(\mu-n)}   \\
&\displaystyle + {1\over {2  \Gamma(\mu)}} 
\int_0^{\infty}ds\ e^{\displaystyle -{\alpha L\over 4}(\mu^2+s^2)}
{{Ês\sinh \pi s} \over {\cosh \pi s - \cos \pi \mu}}
\left\vert
\Gamma\left({\mu\over 2}+i {s\over 2}\right)
\right\vert^2
\label{maxmu+2}
\ea
From (\ref{maxmu-2}-\ref{maxmu=2}-\ref{maxmu+2}), one easily recovers the 
corresponding asymptotic results of (\ref{KT}), which were obtained 
in \cite{Kaw93} by a quite different method,
relying on the computations made in \cite{Yor92a}.
The presence of discrete terms for $\mu >2$ again explains the transition at $\mu=2$
of the asymptotic behavior.

\subsection{ Some relations between $E\bigg( \ln  Z_L^{(\mu) } \bigg)$
and mean inverse $E\bigg( { 1 \over {  Z_L^{(\nu)} } } \bigg)$}

The first negative moment can be obtained
 from the generating function (\ref{genmu}) written above (\cite{Mon94})
\ba
&E\bigg( { 1 \over {  Z_L^{(\mu)}} } \bigg) =
\displaystyle \int_0^{\infty} {dp } \ 
 E\left( e^{\displaystyle - p Z_L^{(\mu)}} \right)  \\
& = \displaystyle \alpha \sum_{0\leq n<{\mu\over 2}} { (\mu-2n) } \ e^{  \displaystyle
-\alpha L n(\mu-n)}
+{\alpha  \over 2 } \int_0^{\infty} ds\ 
 { {s \sinh\pi s } \over {\cosh \pi s -\cos\pi\mu}}\ 
e^{  \displaystyle -{\alpha L\over 4}(\mu^2+s^2)}  
\ea
The previous explicit expressions therefore lead to the
 very simple identity for any $\mu \geq 0$
\be
{\partial \over { \partial L}} E\bigg( \ln  Z_L^{(\mu) } \bigg) = 
E\bigg( { 1 \over {  Z_L^{(\mu)}} } \bigg) - \alpha \mu
\label{idln1z}
\ee
Can it be derived directly using only basic properties of Brownian motion ?
Trying to do so gives in fact two other identities of the same kind,
but not (\ref{idln1z}).
The first one relates the exponential functionals for two opposite drifts
 $(+\mu)$ and $(-\mu)$ 
\be
{\partial \over { \partial L}} E\bigg( \ln  Z_L^{(\mu) } \bigg) =
E\bigg( {1 \over {  Z_L^{(-\mu)}} }\bigg) 
\ee
This identity can be obtained from a simple reparametrisation $x'=L-x$
 in the denominator of the left handside of expression
\be
{\partial \over { \partial L}} E\bigg( \ln \int_0^L dx \ e^{\displaystyle
 -( \alpha \mu x +\sqrt{2 \alpha} B_x )} \bigg) =
 E\left( { { e^{\displaystyle -( \alpha \mu L +\sqrt{2 \alpha} B_L )} } \over { 
\int_0^L dx \ e^{\displaystyle -( \alpha \mu x +\sqrt{2 \alpha} B_x )}}}  \right)
\ee
The second one relates the exponential functionals for two dimensionless drifts
 differing by two 
\be
{\partial \over { \partial L}} E\bigg( \ln  Z_L^{(\mu) } \bigg) =
e^{- \displaystyle \alpha L (\mu-1)} \ E\bigg( {1 \over {  Z_L^{(\mu-2)}}}\bigg) 
\ee
This identity follows from Cameron-Martin or Girsanov relations.
In fact these relations lead to a more general relation between the two characteristic
 functions of the exponential functionals for $(\mu)$ and $(\mu-2)$
\be
{\partial \over { \partial L}} E\bigg( e^{-p  Z_L^{(\mu)}}  \bigg) =
-p \ e^{- \displaystyle \alpha L (\mu-1)} \ E\bigg( e^{-p  Z_L^{(\mu-2)}}\bigg) 
\ee

\vskip 1 true cm

\section{   Expressions of
 $E \bigg(\ln \big( Z_{L_{\lambda}}^{(\mu)}  \big) \bigg) $
 with an independent exponential length $L_{\lambda}$ }
\label{explength}

It is well known that the laws of additive functionals 
of Brownian motion with drift $\mu$
\be
{\cal A}_t^f \quad \stackrel{\hbox{\scriptsize(def)}}{=} \quad  \int_0^t ds \ f(B_s+\mu s) 
\ee
may be easier to compute when the fixed time $t$ is replaced by 
an independent exponential time $T_{\lambda}$ of parameter $\lambda$
\be
P \bigg( \ T_{\lambda} \in [t,t+dt] \  \bigg) = \lambda e^{-\lambda t} dt
\ee
This is indeed the case for ${\cal A}_t^f= A_t^{(\mu)}$ i.e. $f(x)=e^{-2x}$.
It was shown in \cite{Yor92b}  \cite{Yorbook} that
\be
 A_{T_{\lambda}}^{(\mu)} \quad \stackrel{\hbox{\scriptsize(law)}}{=}
 \quad  {X_{1,a} \over {2 Y_b}}
\qquad \hbox{where} \qquad a={ {\sqrt{2 \lambda +\mu^2} -\mu} \over 2} , \qquad
b={ {\sqrt{2 \lambda +\mu^2} +\mu} \over 2},  
\ee  
 $X_{\alpha, \beta}$ denotes a Beta-variable with parameters $(\alpha, \beta)$
\be
P\bigg ( \ X_{\alpha, \beta} \in [x,x+dx] \ \bigg) = 
{ {x^{\alpha-1} (1-x)^{\beta-1}} \over
B(\alpha,\beta) }dx  \qquad (0<x<1)
\ee
and 
$Y_{\gamma}$ denotes a Gamma-variable of parameter $(\gamma)$ (\ref{gamma}).
Consequently, one has
\be
E \bigg(  \left(  A_{T_{\lambda} }^{(\mu) } \right)^n  \bigg) =
{ { \Gamma(1+n)  \Gamma(1+a) \Gamma(b-n)} \over { 2^n \Gamma(1+a+n) \Gamma(b)}Ê}
\ee
and
\be
-E \bigg( \ln  A_{T_{\lambda} }^{(\mu) }  \bigg)  =C+ \ln 2+ \psi(1+a)+\psi(b)
\label{lnlambda}
\ee
where $C=-\Gamma'(1)$ is Euler's constant, and $\psi(x) = { \Gamma'(x) \over \Gamma(x) }$.
Classical integral representation of the function $\psi$ 
allows to invert the Laplace transform in $\lambda$ implicit in (\ref{lnlambda}),
hence to recover $E \bigg( \ln  A_t^{(\mu) }  \bigg)$; however the formulae we have obtained in
this way are not simple.

To transpose the result (\ref{lnlambda}) for the partition function
$Z_{L_{\lambda} }^{(\mu) } $ 
describing the case where the length of the 
disordered sample is exponentially distributed, one only needs
to use the identity derived from (\ref{ZAlaw})
 \be
Z_{L_{\lambda} }^{(\mu) } \quad \stackrel{\hbox{\scriptsize(law)}}{=} \quad  
{2 \over \alpha}\  A_{T_{{2 \over \alpha} \lambda} }^{(\mu)}
\ee

\vskip 1 true cm

\centerline{ \bf Appendix : A simple proof of Bougerol's identity}

\vskip 0.5 true cm

As Bougerol's identity (\ref{Bougerol}) may appear quite mysterious at first sight,
we find useful to reproduce here a simple proof
 of this identity due to L. Alili and D. Dufresne.
We refer the reader to \cite{Ali95} and \cite{Ali96} for further details and
possible generalizations for the case of a non-vanishing drift $\mu \neq 0$.

Consider the Markov process
\be
X_t \ =e^{B_t} \ \int_0^t \ e^{-B_s} \ d \gamma_s
\ee
where $B_t$ and $\gamma_t$ are two independent Brownian motions.
 It\^o formula yields the stochastic differential equation
\be
dX_t \  = {1 \over 2} X_t \  dt + ( X_t \  dB_t+ d \gamma_t)
\ee 
We introduce a new Brownian motion $\beta_t$ by setting 
\be
X_t \  dB_t+ d \gamma_t = \sqrt{X_t^2+1} \ d \beta_t
\ee
from which it follows that
\be
dX_t \ = {1 \over 2} X_t \  dt + \left(X_t^2+1 \right)^{1\over 2} \ d \beta_t
\ee 
The comparison with 
\be
d \big[ \sinh(\beta_t) \big] =
 {1 \over 2} \big[ \sinh(\beta_t) \big] dt +
 \left(\sinh^2(\beta_t) +1 \right)^{1\over 2}  \ d \beta_t
\ee 
shows that
\be
\sinh(\beta_t) \quad \stackrel{\hbox{\scriptsize(law)}}{=} \quad 
X_t \  = e^{B_t} \ \int_0^t \ e^{-B_s} \ d \gamma_s
\ee
The use of scaling properties of Brownian motion finally gives
\be
\sinh(\beta_t) \quad \stackrel{\hbox{\scriptsize(law)}}{=} \quad 
\hat{\gamma}_{A_t^{(0)}} \qquad \hbox{with}
 \quad A_t^{(0)}= \int_0^t \ e^{2 B_s} \ ds
\ee
where $\hat{\gamma}$ denotes a Brownian motion, which is independent of $B_s$.

\vskip 2 true cm 

{\bf Acknowledgments}

We thank J.P. Bouchaud for interesting discussions at the beginning
 of this work, and B. Duplantier for a careful reading of the manuscript. 

\vskip 2 true cm

\end{document}